\documentstyle[11pt,psfig,aas2pp4]{article}

\def\cf{{\em cf.\/}}
\def\ie{{\it i.e.\/}}
\def\secondip{\hbox{\rlap{\hbox{.}}\hbox{$''$}}}
\def\primip{\hbox{\rlap{\hbox{.}}\hbox{$'$}}}
\def\gradip{\hbox{\rlap{\hbox{.}}\raise 5.truept \hbox{{\small $\circ$}}}}

\catcode`\@=11
\def\gsim{\ifmmode{\mathrel{\mathpalette\@versim>}}
    \else{$\mathrel{\mathpalette\@versim>}$}\fi}
\def\lsim{\ifmmode{\mathrel{\mathpalette\@versim<}}
    \else{$\mathrel{\mathpalette\@versim<}$}\fi}
\def\@versim#1#2{\lower 2.9truept \vbox{\baselineskip 0pt \lineskip
    0.5truept \ialign{$\m@th#1\hfil##\hfil$\crcr#2\crcr\sim\crcr}}}
\catcode`\@=12

\begin{document}
\title{The White Dwarf Distance to the Globular Cluster 47~Tucanae and its
	Age\altaffilmark{1}}
\author{
M. Zoccali\altaffilmark{2}, 
A. Renzini\altaffilmark{2}, 
S. Ortolani\altaffilmark{3}, 
A. Bragaglia\altaffilmark{4}, 
R. Bohlin\altaffilmark{5}, 
E. Carretta\altaffilmark{6}, 
F. R. Ferraro\altaffilmark{4}, 
R. Gilmozzi\altaffilmark{2}, 
J. B. Holberg\altaffilmark{7}, 
G. Marconi\altaffilmark{8}, 
R. M. Rich\altaffilmark{9}, 
F. Wesemael\altaffilmark{10}}

\altaffiltext{1}{Based on observations with the NASA/ESA {\it Hubble Space 
Telescope}, obtained at the Space Telescope Science Institute, which is 
operated by AURA, Inc., under NASA contract NAS5-26555.}

\altaffiltext{2}{European Southern Observatory, Karl-Schwarzschild 2, D-85748 
Garching b. M\"{u}nchen, Germany; mzoccali@eso.org, arenzini@eso.org, 
rgilmozz@eso.org}

\altaffiltext{3}{Dipartimento di Astronomia, Universit\`a di Padova, vicolo 
dell'Osservatorio 5, I-35122 -- Padova -- Italy; ortolani@pd.astro.it}

\altaffiltext{4}{Osservatorio Astronomico, via Ranzani 1, I-40127 Bologna, Italy;
angela@bo.astro.it, ferraro@bo.astro.it}

\altaffiltext{5}{Space Telescope Science Institute, 3700 San Martin Drive, 
Baltimore, Maryland 20771; bohlin@stsci.edu}

\altaffiltext{6}{Osservatorio Astronomico, vicolo dell'Osservatorio 5, I-35122 -- 
Padova -- Italy; carretta@pd.astro.it}

\altaffiltext{7}{Lunar and Planetary Laboratory, University of Arizona, Tucson, 
AZ~85721, USA; holberg@argus.lpl.arizona.edu}

\altaffiltext{8}{Osservatorio Astronomico di Roma, Roma, Italy; gmarconi@eso.org}

\altaffiltext{9}{Department of Physics and Astronomy, Division of Astronomy and 
Astrophysics, University of California, Los Angeles, CA 90095-1562; 
rmr@astro.ucla.edu}

\altaffiltext{10}{D\'epartement de Physique, Universit\'e de Montreal, CP6128, 
Succ.\ Centre-Ville, H3C 3J7, Montr\'eal, Qu\'ebec, Canada; wesemael@astro.umontreal.ca}


\begin{abstract}

We present a new determination of the distance (and age) of the
Galactic globular cluster 47~Tucanae (NGC~104) based on the fit of its
white dwarf (WD) cooling sequence with the empirical fiducial sequence
of local WD with known trigonometric parallax, following the method
described in Renzini et al.\ (1996).  Both the cluster and the local
WDs were imaged with HST+WFPC2 using the same instrument setup. We
obtained an apparent distance modulus of $(m-M)_V=13.27\pm0.14$
consistent with previous ground-based determinations and shorter than
that found using {\it HIPPARCOS} subdwarfs. Coupling our distance
determination with a new measure of the apparent magnitude of the main
sequence turnoff, based on our HST data, we derive an age of
$13\pm2.5$ Gyr.

\end{abstract}


\section{Introduction}

Galactic globular clusters (GCs) are the oldest systems for which ages
can be accurately measured using the stellar evolutionary clock.  As
such, they constrain the age of the Galaxy, hence of the Universe, and
by extension the choice of cosmological parameters. The only
observable quantity that can be used to measure the {\it absolute} age
of a GC is the absolute magnitude of the main sequence turnoff
($M_{\rm V}^{\rm TO}$). Other observables are also sensitive to age,
e.g. the horizontal branch (HB) morphology or the turnoff (TO) colors,
but their use is dangerous since in addition to age they are sensitive
to other parameters such as mass loss and convective efficiency,
respectively.

The TO luminosity of a GC is derived from the observed magnitude of TO
stars, correcting for interstellar extinction, and applying the
distance modulus. Simple calculations show that the age ($t$) error
budget is indeed dominated by the uncertainty in GC distances, with
the simple rule $\delta t/t=\delta$mod (Renzini 1991), with e.g., an
uncertainty of 0.25 mag in distance modulus implying a $\sim 25\%$
uncertainty in age. Correspondingly, efforts at improving the accuracy
of GC age determinations have focused on the distance issue.

Although the direct measurement of the parallax of several (perhaps
all) GCs is on the horizon (SIM, GAIA), for the time being we are
still bound to use indirect methods, based on {\it standard
candles}. In most recent years local subdwarfs have been widely used as
standard candles, especially after trigonometric parallax data from the
{\sl HIPPARCOS} mission have become publicly available.

Indeed, {\sl HIPPARCOS} has produced a large sample of local subdwarfs
with accurate parallaxes, and the general perception has been that a
sizable reduction of GC ages using the subdwarf method (Reid 1997,
1998; Gratton et al, 1997) was due to the new set of subdwarf
parallaxes. Actually, the situation is somewhat more intricate. For
subdwarfs in common with previous, ground-based parallaxes the new
{\sl HIPPARCOS} measurements give distance moduli that on average are
just 0.1 mag larger (see Fig. 2 in Reid 1997). This corresponds to a
reduction in age of only $\sim 10\%$, while claimed increases in GC
distance moduli were $\sim 0.3$ mag, and the corresponding reductions
over previous age determinations were up to $\sim 30\%$ (Reid 1997;
Gratton et al. 1997).  Therefore, {\sl HIPPARCOS} parallaxes alone
account for only $\sim 1/3$ of the claimed reductions, much of the
effect coming instead from a variety of cumulative effects, from
including subdwarfs and evolved stars with relatively uncertain
parallax measurements (Reid 1997), to adopting a different metallicity
scale (Gratton et al. 1997). Systematic differences of 1-2 Gyr are
also introduced by the use of different theoretical $M_{\rm V}^{\rm
TO}-$age relations.  
Moreover, contrasting results for M~92 (Pont et al. 1998, Reid 1997
and Carretta et al. 2000) show how much care is required to properly
deal with the many observational biases and reddening and metallicity
scales. In fact, the fortuitous agreement of the distance moduli 
derived by Pont et al.\ and by Carretta et al.\ is a spurious
result, given by adoption of different reddening values and metallicities
for the field subdwarfs used in the fitting. 
In summary, {\sl HIPPARCOS} GC distances and ages still remain matter 
of debate.

Every method of distance determination presents advantages and
disadvantages.  For example, GC distances derived from the subdwarf
method require the metallicity of the cluster and of each subdwarf to
be specified, which makes the method prone to possible systematic
biases of the dwarfs vs. giants metallicity scales (see e.g., King et
al.\ 1998).  Thus, a mismatch of $\sim 0.25$ dex in the subdwarfs
vs. GC metallicity scales results in a $\sim 0.25$ mag error in
distance and in a $25\%$ error in age (Renzini 1991).  For these
reasons, we believe that a variety of methods should be used to derive
GC distances, as discrepancies among the results should help to
identify possible systematic biases.

As an alternative to the subdwarf method, Renzini et al. (1996,
hereafter Paper I) used the DA white dwarf cooling curve as a fiducial
to measure the distance to one of the nearest GCs, NGC 6752, obtaining
a true modulus of 13.05 mag and an age of 14.5 Gyr (including He
diffusion) and 15.5 Gyr (not including He diffusion).  In this paper,
we use the white dwarf (WD) method to derive the distance modulus to
47~Tuc, a globular cluster that is roughly a factor of ten more metal
rich than NGC 6752 ([Fe/H]=$-0.70\pm 0.03$; Carretta \& Gratton 1997).
In extending our approach to this higher metallicity, we will be able
to determine whether there is an age-metallicity relationship among the
Galactic GCs.  The similarity of the Color-Magnitude diagram (CMD)
locus of 47 Tuc to that of bulge globulars NGC 6528 and NGC 6553
(Ortolani et al. 1995) will constrain the ages of bulge clusters of
near solar metallicity, as well as of the field population of the
Galactic bulge.  In principle, the large difference in metallicity
between NGC 6752 and 47~Tuc should be useful in giving an independent
constraint on the relationship between the $V$ magnitude of the
horizontal branch and [Fe/H].  Unfortunately, the horizontal branch of
NGC 6752 is nearly vertical, so its $V$ magnitude is not so well
defined.  On the other hand, 47~Tuc has only one peculiar RR-Lyrae star,
but does have a well defined red HB at $V=14.10$ (Kaluzny et
al. 1997). These two clusters were selected for the
application of the WD method for being the least reddened ones among
the nearest GCs.


\section{Observations and Data Analysis}
\label{sec2}

A field located $6\primip5$ West of the center of 47~Tuc was observed
during Cycle-5 with the HST-WFPC2 camera through the filters F336W,
F555W, and F814W.  Short exposures with the same instrument setup +
long F439W band frames were taken during Cycle-7. Finally, some extra
F814W frames of the same field, acquired for astrometric purposes,
where kindly provided by I. King. All in all, the total exposure time
was 26300s, 13800, 15880 and 21949 seconds in F336W, F439W, F555W and
F814W, respectively. Most of the observations were cosmic-ray split.
For the local calibrators, we re-measured the magnitudes of the four
DA WDs used in Paper~I. In order to achieve a better definition of the
local WD sequence, four new local WDs (2 DA + 2 DB) were observed in
the same filters.  One more star (WD1647+591) was observed but not
used because its location in the CMD is too red for its mass (0.69
M$_\odot$). We conclude that either it is a binary, or that there must
be an error in the mass determination and/or in the parallax.

All the images were de--biased, dark corrected and flatfielded through
the standard HST pipeline. Following Silbermann et al. (1996), we have
masked out the vignetted region, the saturated and bad pixels and
columns using a vignetting mask created by P.B. Stetson, together with
the appropriate data quality file for each frame. 

\subsection{Cluster data}

The photometric reduction of the frames in the field of 47~Tuc was
carried out using the DAOPHOT~II/ALLFRAME package (Stetson 1987,
1994).  Preliminary photometry was performed on each single frame in
order to obtain an approximate list of stars for each of them. The
coordinates of the common stars were used for an accurate spatial
match among the different frames.  With the correct coordinate
tranformations, we then obtained a single image, which was the median
of all the frames, regardless of the filter. In this way we removed
all the cosmic rays and obtain the highest signal-to-noise image for
star identification.  We ran the DAOPHOT/FIND routine on the median
image and performed PSF fitting photometry on it, in order to obtain
the deepest list of stellar objects, free from spurious
detections. Finally, this list was given as input to ALLFRAME, for the
simultaneous profile fitting photometry of all the individual
frames. The PSF we used were the WFPC2 reference PSFs extracted by
P.B.Stetson (1995) from a large set of uncrowded and unsaturated
images.

For a consistent comparison between the 47~Tuc and the local WD stars,
all the measured instrumental magnitudes were scaled to a 1s
exposure. Aperture corrections for the cluster data were measured from
a few isolated stars in each chip and filter. They were applied to the
cluster stars in order to convert the PSF magnitudes to the same zero
point as the local WDs, measured with an aperture of $0\secondip5$
radius.  The stars observed with each of the WF chips were then put
together, after removing the small chip--to--chip offset (Dolphin 2000). 

It is now well known that the WFPC2 detector is affected by a charge
transfer efficiency (CTE) problem: a loss of charge during the CCD
transfer phase. The amount of the loss depends on the position within
the chip, the background counts, and has worsened with time since the
installation of the WFPC2 camera.  In the last few years various
estimates of the size of this effect have been made (e.g., Whitmore \&
Heyer 1997; Stetson 1998; Whitmore, Heyer \& Casertano 1999) that was
sometimes mis-interpreted as a ``long vs short anomaly'' (Casertano \&
Mutchler 1998). However only very recently Dolphin (2000) provided a
formula that takes into account all the dependencies of the charge
loss, including its variation with the frame background (and therefore
with the exposure time) and its increase with the epoch of the
observation.  The CTE problem is critical for our method, since it
relies mainly on the photometry of faint stars in low background (the
background being especially low in the F336W frames). For this reason,
we have paid special attention in applying the CTE correction (Dolphin
2000) and in veryfing that it gives sound results.
The size of the correction for a cluster WD in F336W ranges from a few
hundredth up to ~0.3 magnitudes, depending on its magnitude and
position in the chip.  However, the brightest cluster WDs have been
corrected by less than 0.05 mag, and therefore if a residual CTE
problem were present, a large difference between the slope of the
cluster and the local WD sequence would be present. The fact that no
such difference is apparent (Fig.~\ref{fit}) makes us confident about
the adopted correction.  Finally, a further correction, of the order
of a few hundredth of a magnitude, was applied to the F336W magnitudes
to account for the time dependence of the UV throughput (Baggett \&
Gonzaga 1998).

\begin{figure*}
\vskip 0.4cm
\centerline{\psfig{file=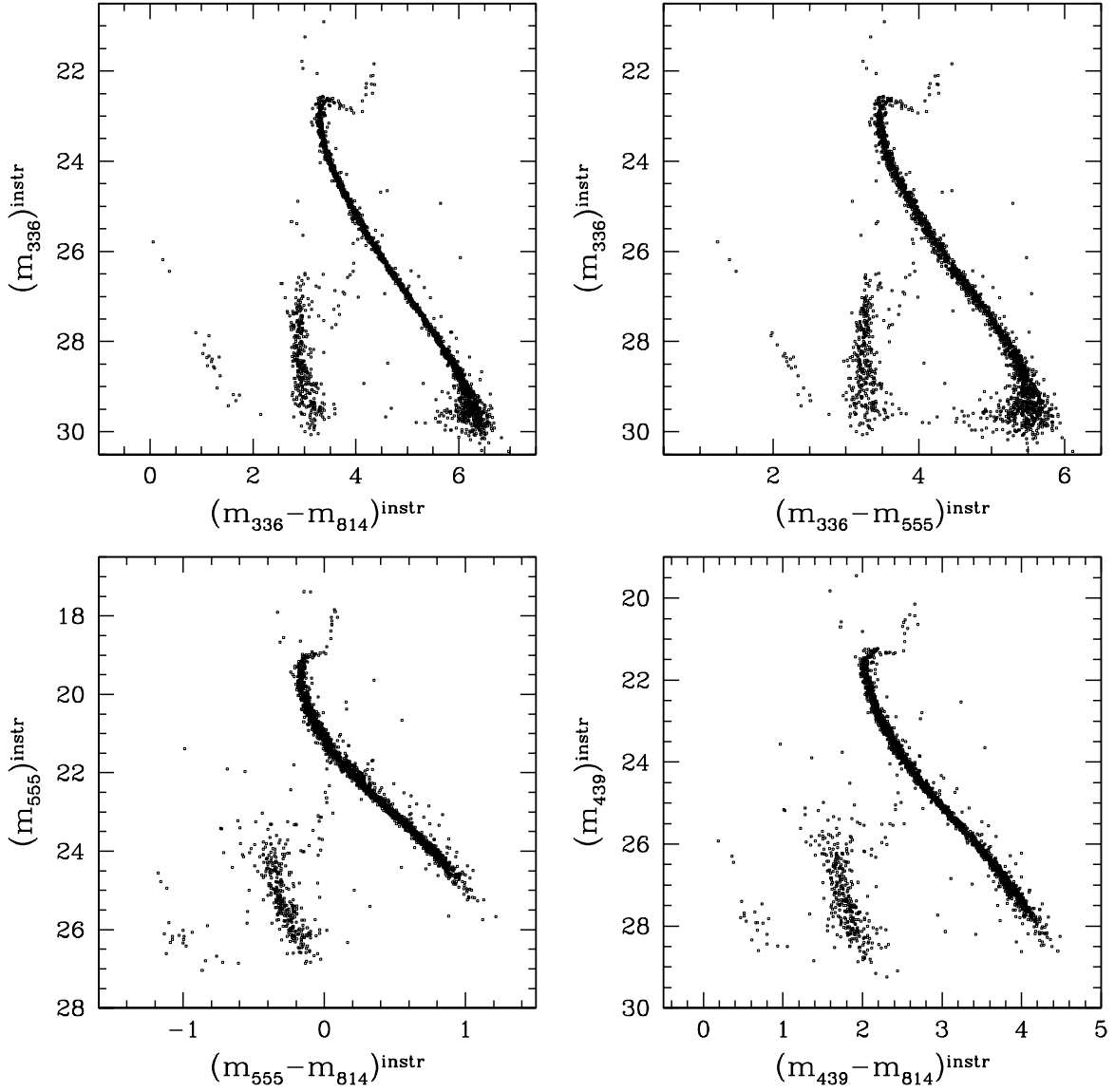,angle=0,width=17truecm}}
\caption[]{The instrumental CMD of the observed 47~Tuc field. Three 
main branches are clearly distinguishable: the cluster MS on the 
right, the SMC MS and SGB in the middle, and the cluster white 
dwarf sequence on the left.}
\label{cmd}
\end{figure*}

\begin{figure*}
\vskip 0.4cm
\centerline{\psfig{file=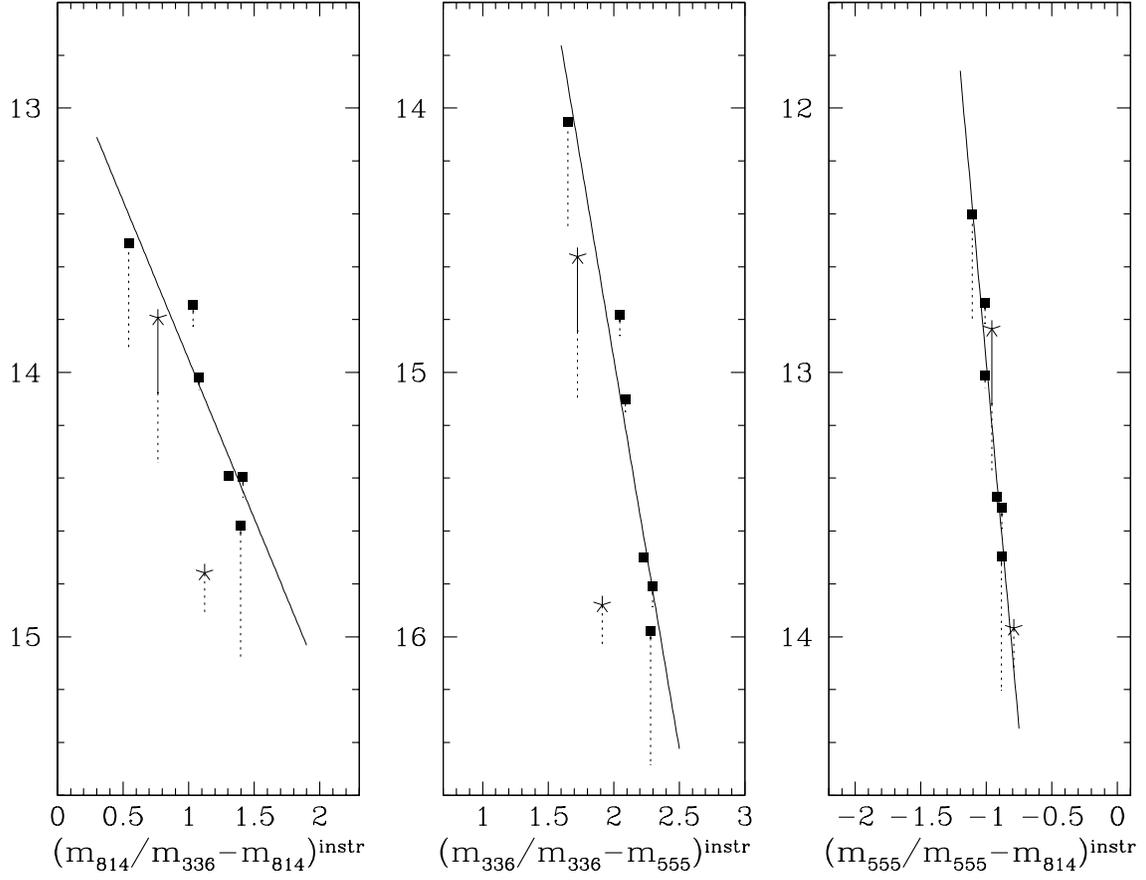,angle=-90,width=17truecm}}
\caption[]{Instrumental CMDs of the local WDs used as calibrators. DA WDs 
are plotted as filled squares, DBs are the asterisks. A least square fit to
the DA WDs is shown as a solid line. The vertical dotted lines show 
the size of the magnitude correction applied to the WDs whose masses differ 
from 0.53, while the solid vertical line is the size of the Lutz-Kelker
correction. }
\label{local}
\end{figure*}

The instrumental CMDs containing the stars from the three WF, in the
four color planes, are shown in Fig.~\ref{cmd}. The photometry from
the PC chip was not used because no WDs were detected in this field.
In what follows we will use the notation $m_{336}$, $m_{439}$,
$m_{555}$, $m_{814}$ to indicate the 1s instrumental magnitudes in the
F336W, F439W, F555W and F814 filters, respectively.  Only the stars
detected in $\ge 50\%$ of the frames in each filter were accepted, and
a further selection based on the magnitude errors and the PSF fitting
parameters was made. Fig.~\ref{cmd} shows only the stars ($\sim2300$)
in common to the four diagrams, for this reason the main sequence in
the ($m_{555}$, $m_{555}-m_{814}$) plane is truncated at a magnitude
brighter than the $m_{555}$ limit magnitude.  A WD sequence is clearly
visible in all four diagrams, although it is better defined in the
($m_{336}, m_{336}-m_{555}$) and ($m_{336}, m_{336}-m_{814}$) planes.
The main sequence and subgiant branch of the Small Magellanic Cloud
old population are also clearly visible at an intermediate location
between the 47~Tuc main sequence and the WD sequence.  A total of 21
cluster WDs were identified. The WD sequence appears very narrow
especially in the two CMDs using the $m_{336}$ magnitudes, which, due
to the larger color baseline, are also the most useful one to
distinguish between DA and DB dwarfs (\cf\ Fig.~\ref{local}).
Therefore, we can conclude that only DA WDs have been sampled in this
47~Tuc field, and therefore only the local WDs of the DA type will be
used as calibrators for the distance determination.

\subsection{Local WD data}

The local WD calibrators were selected for having a trigonometric 
parallax with accuracy better than $10\%$ {\it and} a spectroscopically
determined mass as close as possible to 0.53 $M_\odot$, the predicted
mass for the globular cluster WDs (Renzini \& Fusi Pecci 1988; see
also Fig.~1 in Greggio and Renzini 1990).

Every local WD was imaged by each of the 3 WF chips. Aperture
photometry was performed within a radius of 0.5 arcsec on each image,
the chip--to--chip offset, as determined by Dolphin (2000), was
removed, and then the three magnitudes of each star were averaged.
The instrumental magnitudes were scaled to 1s exposure and corrected
for the decreasing UV throughput ($m_{336}$ only).  The absolute
instrumental magnitudes were then obtained by means of the
trigonometric parallaxes reported in Table~1. Three of our local
calibrating WDs were observed by HIPPARCOS, and their parallaxes,
although slightly different from the ground-based ones, do not show
any systematic zero point shift. These stars are WD0644+375,
WD1327+276, and WD1917-077, and the percent differences
($\pi_{HIP}-\pi_{VA}$) between the parallaxes in the {\it HIPPARCOS}
and the Van Altena et al. (1991) catalog are: $+4\%$, $-9\%$, $-12\%$,
respectively. Note that the latter is a DB WD, so it was not used in
the fit.  In addition, Vauclair et al.\ (1997) showed that the
comparison between the parallaxes the 16 DA WDs in common between the
two catalogs reveals no systematic difference.

The Lutz-Kelker correction, calculated according to the formula given
in Carretta et al.\ (2000), amounts to a few hundredth of a magnitudes
for all the stars except one, the DB WD WD0002+729, for which it is
$\sim 0.3$ magnitudes. However, this star, being of DB type, was not
used for the distance determination.  Table~1 also shows that the
masses $M_{\rm WD}$ of a few of the selected local WDs differ
appreciably from the ideal value, 0.53 $M_\odot$.  In order to obtain
a constant-mass sequence appropriate for the match with the cluster
WDs, the correction $\delta (mag) = 2.4(M_{\rm WD} - 0.53)$ as applied
to the magnitudes of all the local WDs (Wood 1995; Bergeron, Wesemael
\& Beauchamp 1995). Moreover, the WD masses listed in Table~1 were
all (except WD2326+049) originally derived adopting a WD mass-gravity
relation appropriate for WDs without an hydrogen envelope. However, as
extensively discussed in Bragaglia, Renzini \& Bergeron (1995), there
is now general consensus that ``evolutionary'' values for the mass of
the hydrogen envelope should be used, i.e., $\sim 10^{-4} M_\odot$ for
the WD masses of interest here (Napiwotzki, Green \& Saffer, 1999).
This implies that WD masses in Table~1 were underestimated by $\sim
0.04 M_\odot$ (Bragaglia et al.  1995), and therefore such masses have
been increased by this amount in producing the final fiducial WD
sequence used in the distance determination.  The magnitude correction
for each calibrating WD (except WD2326+049 that was obtained with
the correct assumption on the envelope thickness) was then taken as $\delta
(mag) = 2.4(M_{\rm WD}+0.04-0.53)$, where $M_{\rm WD}$ are the masses
in Table~1.

\begin{figure*}
\vskip 0.4cm
\centerline{\psfig{file=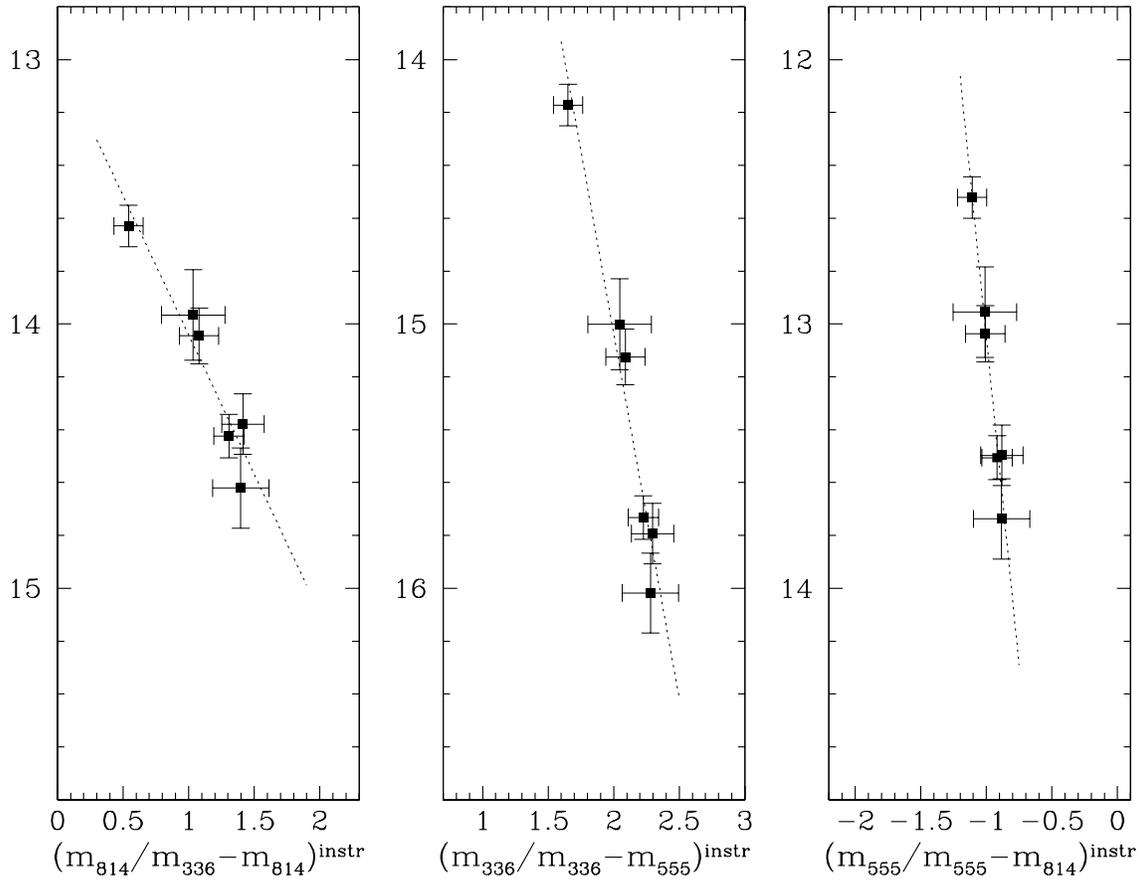,angle=-90,width=17truecm}}
\caption[]{Same as Fig.~\ref{local}, but only for the DA WDs
and with errorbars. The latter include the internal photometric
errors, as well as the uncertainties in the parallax and mass
determinations listed in Table~1.}
\label{localerr}
\end{figure*}

The resulting fiducial WD sequences are shown in Fig.~\ref{local},
including the best fit straight line in each individual CMD. This plot
also shows the size of the Lutz-Kelker (solid vertical lines) and of
the mass corrections (dotted lines) applied to each star.  Note that
the dispersion is very small and that the one reliable DB dwarf is
located well outside of the DA sequence only in the ($m_{336},
m_{336}-m_{555}$) plane. As mentioned above, no star in the 47~Tuc
CMDs matches the location of the DB sequence.  Figure~\ref{localerr}
shows the same CMDs but only for the WDs used for the distance
determination, with the estimated errors being also indicated.

In summary, the WD fiducial cooling sequence relies on the
spectroscopically determined gravities, a theoretical mass-gravity
relation corrected for the finite mass of the hydrogen envelope, and
on the adopted mass of the cluster WDs, $0.53 \pm 0.02 M_\odot$. This
latter assumption is extensively justified in Paper I, using a series
of convergent stellar evolution arguments that we believe are quite
solid. Of course, the direct spectroscopic measure of the mass of
cluster WDs would be quite reassuring. VLT spectroscopic observations
of a sample of four HST selected WDs in the globular cluster NGC~6397
have recently confirmed their nature of WDs of the DA variety (Moehler
et al. 2000), although the low S/N of the relatively short exposure
spectra of these very faint stars did not allow an accurate
determination of the mass.


\subsection{Photometric zero point check}

Globular cluster distance determinations have to rely on standard
candles, at least until the direct measure of trigonometric parallaxes
will be feasible with SIM/GAIA. The results of the current methods of
distance determinations largely rely on the consistency between the
observations of the standard candles and their cluster counterparts.
Therefore, ideally, the two samples should be measured precisely in
the same conditions. However, in most cases this approach is not
viable.  In the present work, although we observed both the local and
the cluster WDs with the same instrument setup, this fact alone does
not guarantee the necessary homogeneity, given that we are comparing
the photometry from very deep exposures of a sample of extremely faint
stars in a crowded field (\ie\ object for which PSF fitting photometry
is needed) with the aperture photometry of a few very bright stars in
short-exposure frames of almost empty fields. Therefore, in order to
verify that no residual photometric zero point difference can affect
our conclusions, we report here the results of some tests we carried
out to check the consistency of our measurements, i.e., the linearity
of WFPC2 over a range of $\sim 10$ magnitudes.

\begin{figure*}
\vskip 0.4cm
\centerline{\psfig{file=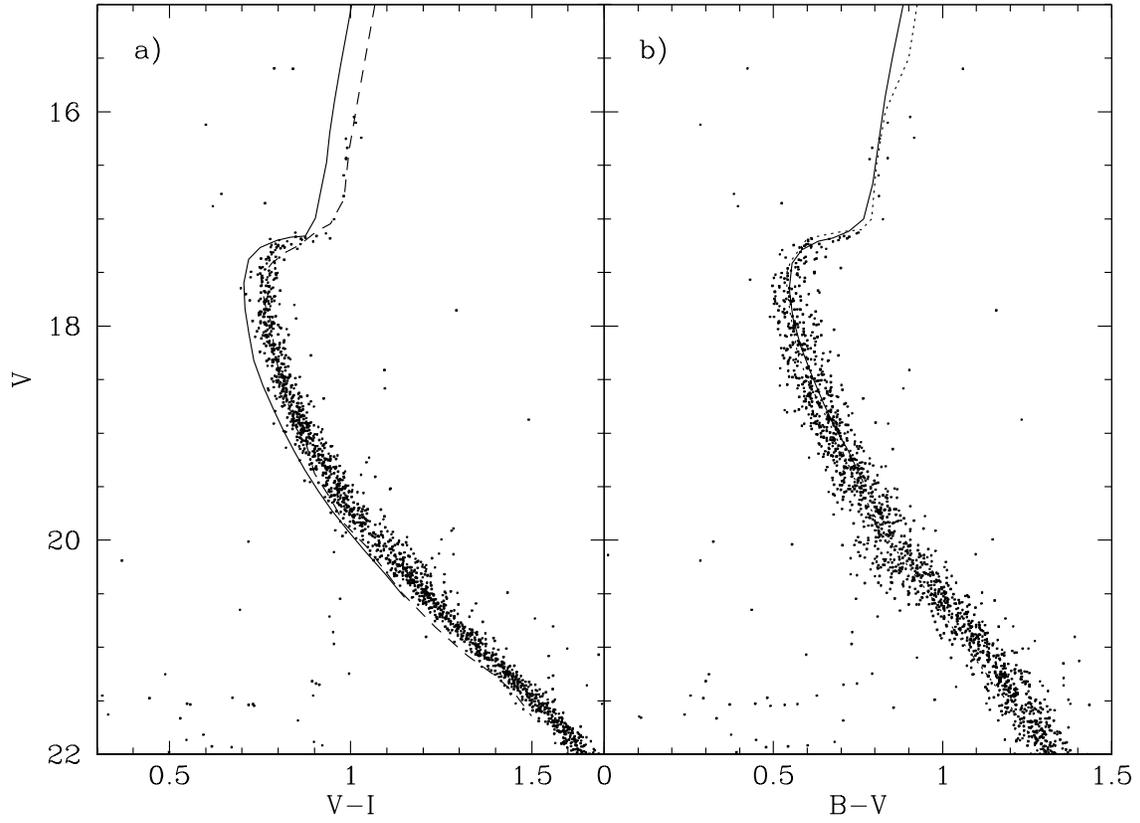,angle=-90,width=17truecm}}
\caption[]{Comparison between our calibrated data and the fiducial
lines of previously published photometries. Left panel: V, V-I
CMD compared with the fiducial lines from Kaluzny et al.\ (1997;
solid) and Ortolani 2000 (dashed). Right panel: B, B-V CMD 
compared with Kaluzny et al.\ (1997; solid), and Hesser et al.\ (1987; 
dotted).} 
\label{calib}
\end{figure*}

The first, most obvious  test of the instrumental magnitudes is their
calibration to the Landolt system (following Holtzmann et al.\ 1995
and Dolphin 2000), and the comparison of the measured calibrated
magnitudes with other photometry available in the literature.  We
emphasize again that we will use only instrumental magnitudes to
estimate the distance of 47~Tuc. This test on the calibrated CMD has
the sole purpose of checking for the presence of some systematic
errors in the instrumental magnitudes, that of course would show up in
the calibrated ones. Since no other photometry is available for our
field of 47~Tuc, it is possible only to compare our CMD with the
fiducial lines of the previous photometry.  Fig.~\ref{calib}a shows
the comparison between our calibrated $(V,V-I)$ diagram and the
fiducial lines of the photometries by Kaluzny et al. (1997; solid) and
Ortolani (private communication; dashed). There appears to be
excellent agreement between Ortolani's photometry and this work, while
Kaluzny's data seem to be systematically bluer by $\sim 0.07$
magnitudes.  Unfortunately we have no means of establishing which one
of the two sets of ground-based photometry is the most correct.
Fig.~\ref{calib}b shows the comparison among our $(V,B-V)$ CMD and
the fiducial sequences by Kaluzny et al. (1997; solid) and Hesser et
al.\ (1987, dotted). In this case there is excellent agreement.
Figure~\ref{calib} makes us confident about the zero point of our
$m_{439}$, $m_{555}$ and $m_{814}$ magnitudes.  In particular, the
above test is also a check of systematic errors in the aperture
corrections, that would affect our distance determination.
 
\begin{figure*}
\vskip 0.4cm
\centerline{\psfig{file=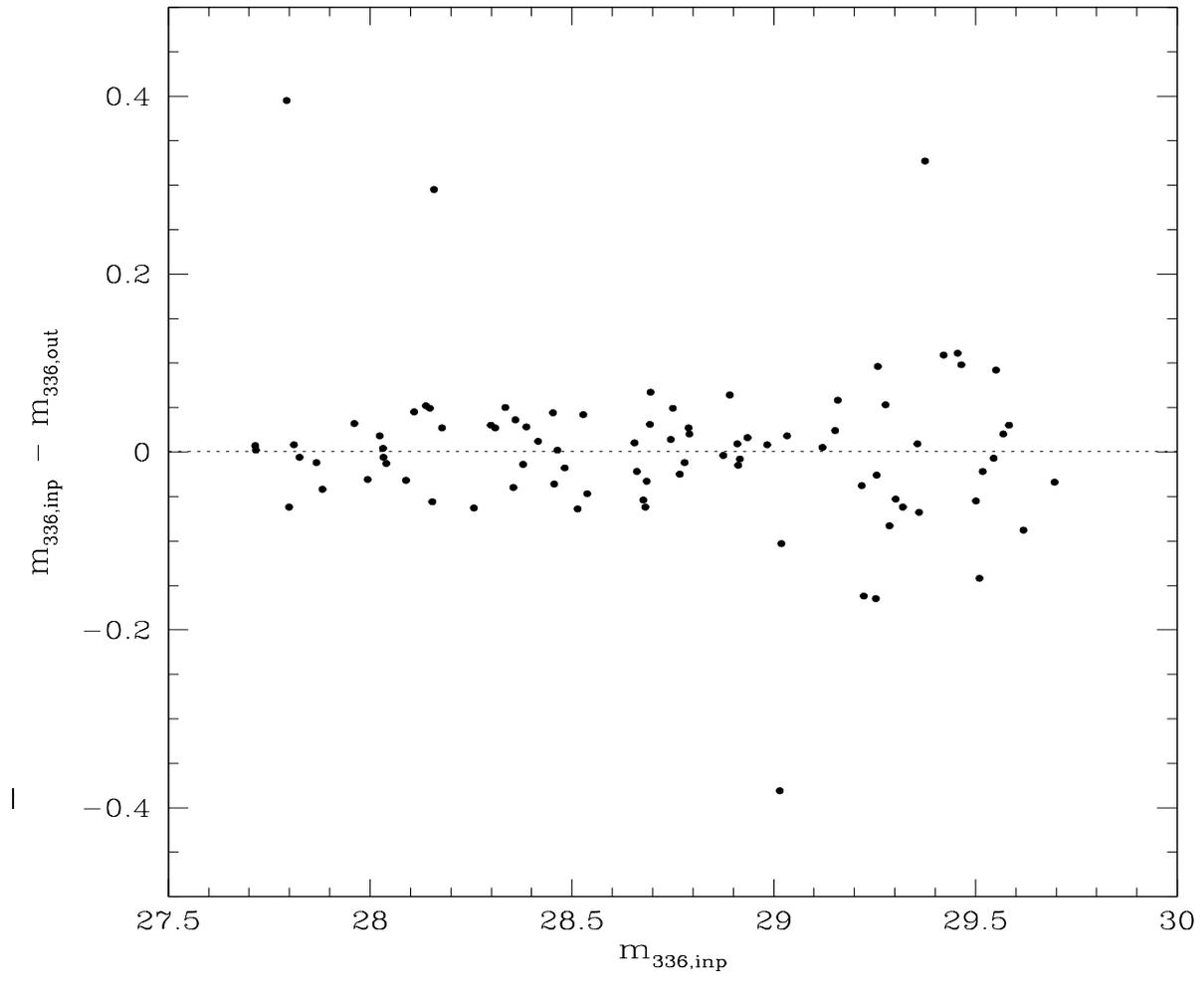,angle=0,height=15cm,width=17cm}}
\caption[]{Artificial star test in the F336W band. The difference between the
input and output magnitudes are plotted as a function of the input magnitude.}
\label{crowd}
\end{figure*}

We did not calibrate the $m_{336}$ magnitudes because the F336W
filter bandpass differs significantly from the Johnson $U$; and,
therefore, the calibration is unreliable, especially for the
hottest/coldest stars.  However, in order to check the zero point of
the $m_{336}$ instrumental magnitudes, we performed an artificial star
test on all the F336W frames. We did this experiments on the F336W
frames only because our distance mainly relies on the $m_{336}$
magnitudes (due to the fact that the WD sequence in this band is less
steep, better defined and well separated from the SMC main sequence)
although the F336W S/N for the faintest WDs was not excellent.  100
artificial stars were added, in the same position, on each single
frame, with magnitudes in the range of the cluster WDs.  Complete
DAOPHOTII/ALLFRAME reduction was then carried out on these frames with
the same algorithm used for the original images, but manually adding
these artificial stars to the star list.  The output magnitudes of the
artificial stars were then compared with the input
ones. Fig.~\ref{crowd} shows this comparison: the systematic error in
the output magnitudes is consistent with zero. For this reason we can
exclude the possibility of a displacement of the cluster WD sequence
in the ($m_{336}, m_{336}-m_{814}$) or ($m_{336}, m_{336}-m_{555}$)
diagrams, due to the migration of the magnitudes towards brighter
values, as a consequence of crowding (Stetson \& Harris, 1988).  Of
course, as all the artificial star tests, this uses model PSFs that
could be different from the real ones.  Still, the fact that the mean
difference between input and output magnitude is consistent with zero
gives a good indication that no {\it significant} migration is present
in the $m_{336}$ photometry of the faint stars.  Note that the number
of stars in Fig.~\ref{crowd} is not related to the completeness of our
photometry because the artificial stars were not identified by the
same finding algorithm used for the original ones.


\section{The distance to 47~Tuc}

Once the cooling sequences for the cluster and the local WDs are
defined, the distance modulus is the vertical shift that makes the
cluster sequence overlap the local one.  We adopt a reddening
$E(B-V)=0.055$, which is the mean value of those reported by Zinn
(1980), Reed et al. (1988) and the Str\"omgren catalogue (Hauck \&
Mermilliod 1990). From this value the correspondent interstellar
absorption $A_\lambda$ was derived for the three bands, according to
Table 12 in Holtzmann et al. (1995), and subtracted them from our
magnitudes. Then we shifted the cluster WDs to match the local ones.
In doing so, we adopted for the WD sequence in each cluster CMD the
same slope determined for the local calibrators, leaving only the
zero-point (i.e., the cluster distance modulus) as the adjustable
parameter.

\begin{figure*}
\vskip 0.4cm
\centerline{\psfig{file=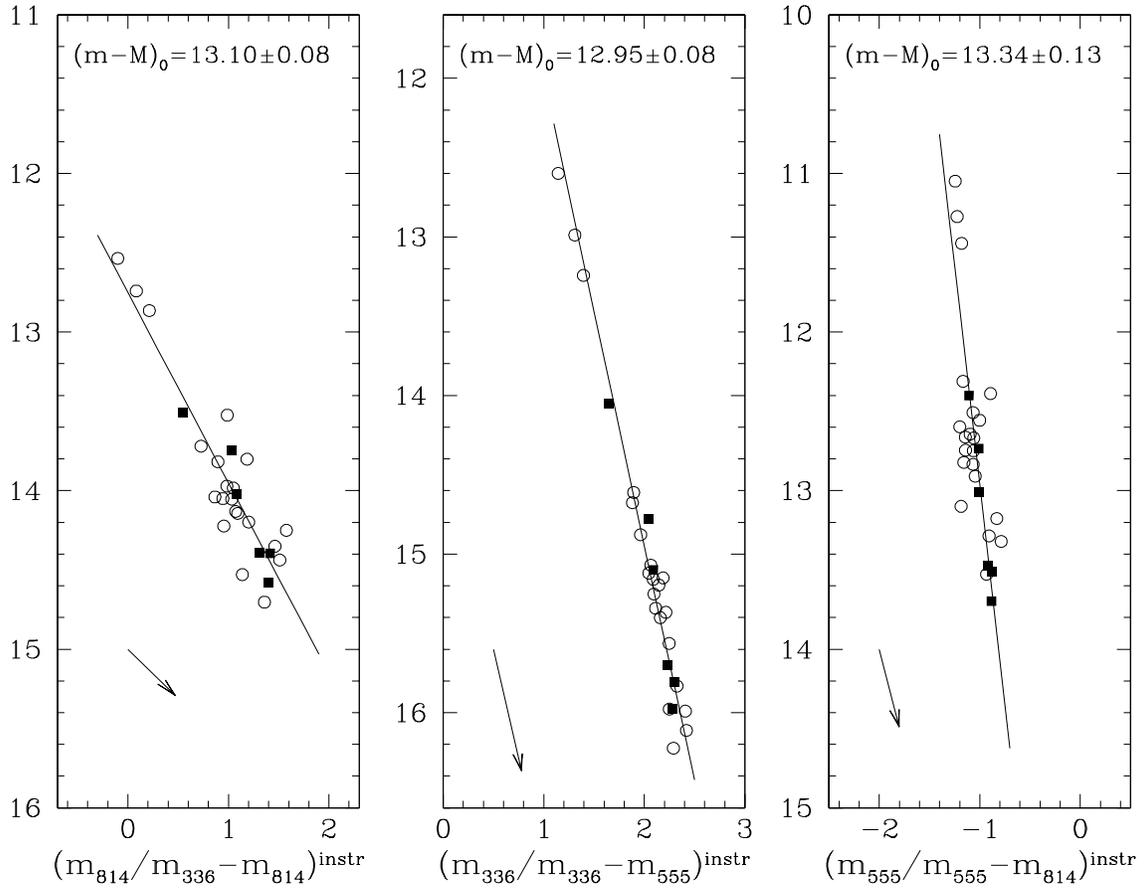,angle=-90,width=17truecm}}
\caption[]{Match between the 21 WDs identified in 47~Tuc (open circles) 
and the local DA WDs (filled squares). The distance moduli required
to match the two sequences in the three planes are shown in the labels,
together with their formal error. The arrow on the lower left corner shows 
the direction of the reddening vector, whose size has been increased by a 
factor of three in order to make it more visible. 
The dotted line in the central and right panel shows the displacement of 
the cluster WD sequence if the mean distance modulus were adopted.
}
\label{fit}
\end{figure*}

Fig.~\ref{fit} shows the match between the two WD sequences in the
three CMDs ($m_{814}, m_{336}-m_{814}$), ($m_{336}, m_{336}-m_{555}$)
and ($m_{555}, m_{555}-m_{814}$).  The distance modulus required to
match the sequences in each panel is shown by the label. The quoted
errors are the internal ones, including the magnitude uncertainties
both in the local and cluster WDs, and the errors in the adopted
parallaxes.  The first two panels allow a more accurate distance
determination, either because the WD sequence is less steep (left
panel), or because of the smaller dispersion around the fiducial
sequence (central panel).  Note that the scale of both axes has been
kept the same in the three panels in order to show the difference in
the sequence slopes.  The arrow in the lower left corner of each panel
shows the {\it direction} of the reddening vector, whose size has been
increased by a factor of three in order to make it more
visible. Depending on the difference in slope between the reddening
vector and the WD sequence, a fixed uncertainty in the cluster
reddening corresponds to a different error in the distance modulus for
each panel. In fact, with an uncertainty in the cluster reddening
$\Delta E(B-V)=\pm0.02$, the corresponding $\Delta (m-M)_0$ are
$\mp0.03$, $\mp0.01$ and $\mp0.07$ in the left, central and right
panels of Fig.~\ref{fit}, respectively. The very small variation of
the distance modulus derived from the $(m_{814},m_{336}-m_{814})$ and
the $(m_{336},m_{336}-m_{555})$ CMDs is due to the fact that in these
planes the WD sequence happens to have a slope very similar to that of
the reddening vector (see arrows in Fig.~\ref{fit}).  Therefore a
change in the adopted reddening shifts the sequence almost parallel to
itself. Adding the above quoted systematic uncertainties to the formal
errors of the fit, and averaging the three results weighted by the
inverse of their errors, we obtain a mean distance modulus of
$(m-M)_0=13.09\pm0.13$.

A further source of systematic error in this value comes from the
adopted mean mass of the cluster WDs: $M_{WD}=0.53\pm0.02$.  According
to the relation quoted above $\delta (mag) = 2.4(M_{\rm WD}-0.53)$ an
uncertainty of 0.02 in mass implies a 0.05 systematic uncertainty in
the distance modulus.  Our best estimate for the distance modulus of
47~Tuc is therefore $(m-M)_0=13.09\pm0.14$. Combining this value with
the adopted reddening $E(B-V)=0.055\pm0.02$ and the extinction law
$A_V=3.2 E(B-V)$ the apparent distance modulus is
$(m-M)_V=13.27\pm0.14$.

In summary, a formula that explicitly shows the dependence of
the apparent distance modulus upon the two main assumptions, namely
cluster WD masses and reddening, can be expressed as follows:

{\footnotesize
\[
(m-M)_V=13.09+3.2\times E(B-V)-2\times[E(B-V)-0.055] 
\]
\[
-2.4\times(M_{\rm WD}^{\rm cl}-0.53)+2.4\times(<M_{\rm WD}>-0.594)
\]
}

\noindent
where $<M_{\rm WD}>$ is the true mean mass of the sample of local WDs,
and $M_{\rm WD}^{\rm cl}$ the true mass of the cluster WDs, and 0.594
is the mean mass of the sampled local DA WDs given in Table~1, including
the correction for the thick hydrogen envelope.

For consistency with Paper I, in our analysis we adopted the
trigonometric parallaxes from the Van Altena et al. (1991) catalog. A
new version of the same catalog, including some more stars and a new
weighting criterion for the various measurements has been released more
recently (Van Altena et al. 1995). Due to the new weighting system,
the parallaxes of two of the local DA WDs, namely WD0644+729 and WD2126+734,
increased by 6$\%$ and 9$\%$, corresponding to a decrease of 0.12 and
0.18 magnitudes in their distance modulus, respectively.  As a
consequence, had we adopted such parallaxes, the distance modulus
for 47~Tuc would have {\it decreased} by $\sim 0.1$ magnitude.

\begin{figure*}
\vskip 0.4cm
\centerline{\psfig{file=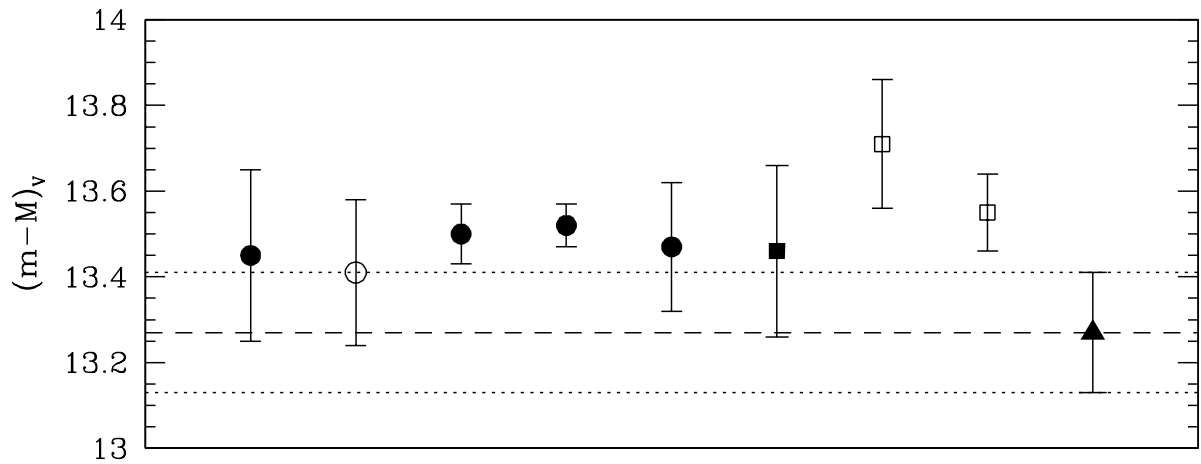,angle=0,width=17truecm}}
\caption[]{Determinations of the distance of 47~Tuc by different
authors, in the same order as Table~2, from left to right. Different
symbols refer to different methods: HB fitting (filled circles),
the unique RR-Lyrae variable (open circle), RGB tip (filled square),
subdwarf fitting (open squares) and this work (filled triangle).}
\label{distance}
\end{figure*}

Table~2 shows a summary of the previous determinations of the distance
of 47~Tuc, using a variety of methods, together with the adopted
reddening. For a direct comparison among the different distances in
Table~2 we reported all to the same reddening $E(B-V)=0.055$, and the
result is plotted in Figure~\ref{distance}, with the corresponding
errorbars.  Our distance determination lies at the lower end of the
distribution.  It is consistent within the errors with most previous
determinations, but it is significantly shorter than the values
reported by Reid (1998) and Carretta et al.\ (2000) using the subdwarf
method based on {\it HIPPARCOS} parallaxes.  The cause of this
discrepancy remains to be understood.

However, were the correct modulus as high as 13.69, i.e., $\sim 0.42$
magnitudes larger than the value indicated by the WD method, then the
inplied mass of the cluster WDs would be as low as $0.53-0.4/2.4=0.36
M_\odot$, much less than indicated by stellar evolution theory.
Alternatively, the mass of the local WDs (via Balmer line fitting)
should have been underestimated by $\sim 0.17 M_\odot$, on average,
which also looks quite implausible.

Still assuming that the WD method is to blame, the discrepancy could
arise from a mismatch between the cluster and the local WD magnitude
scales.  As shown by Fig.~\ref{fit} the distance moduli obtained from
the three CMDs differ appreciably one from another. This can only
arise from a systematic error in the photometry of at least one of the
three bands. Indeed, an error of about $\Delta m_{555}=+0.05$ would be
sufficient to bring the three distance moduli into coincidence, at the
value of $(m-M)_0=13.10$ found from the (obviously unvaried)
($m_{814},m_{336}-m_{814}$) CMD. Similarly, an error of about $\Delta
m_{814}=-0.08$ would make the three distances coincide with the value
of $(m-M)_0=12.95$ found from the middle panel of Fig.~\ref{fit}.  Due
to the still poorly known CTE correction, the presence of such
systematic errors for the very faint cluster WDs cannot be ruled out.
However, this would not help reducing the discrepancy with the
subdwarf distances given by Reid (1998) and Carretta et al.\
(2000). Instead, a systematic error of $+0.2$ in the $m_{336}$
magnitudes only would bring the distance moduli derived from the two
leftmost panels in Fig.~\ref{fit} into coincidence with the value of
$(m-M)_0=13.34$ found from the ($m_{555},m_{555}-m_{814}$) CMD. This
would also solve the discrepancy with the Carretta et al. (2000)
result. However, it would be more likely that there is a systematic
error of the order of $0.05-0.08$ in the $V$ and $I$ instrumental
magnitudes, where the WDs are relatively fainter and the field is more
crowded, rather than an error as large as $0.2$ mag in the $U$ band.

Finally, it is worth noting that were a residual systematic
error the same in {\it all} the magnitudes, then it would have no
effect on the colors, and therefore it would just shift the derived
distance moduli by the same amount (equal to the error itself) in all
the CMDs.


\section{The age of 47~Tuc}

\begin{figure*}
\vskip 0.4cm
\centerline{\psfig{file=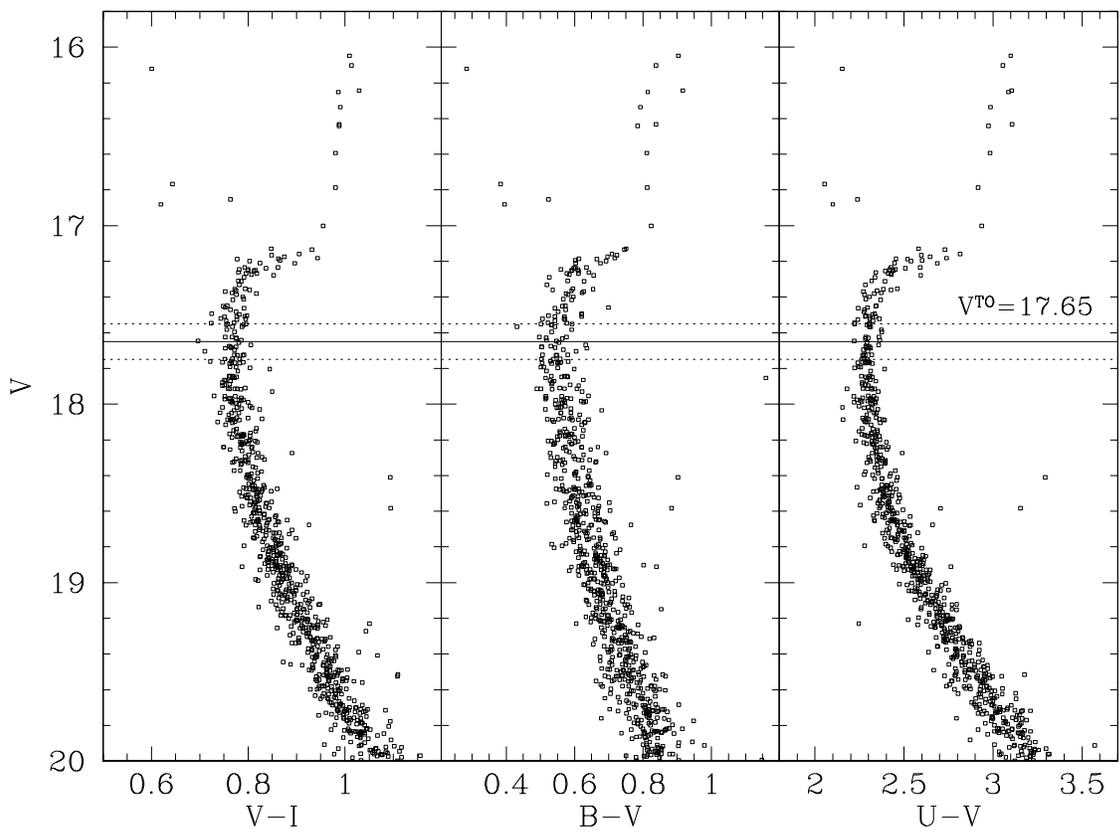,angle=-90,width=17truecm}}
\caption[]{The turnoff region in the calibrated CMD.}
\label{turnoff}
\end{figure*}

The present determination of the distance modulus, combined with the
knowledge of the apparent magnitude of the TO and the global
metallicity, allows us to derive the age of the cluster, via the
turnoff luminosity $M_{\rm V}^{\rm TO}$.  Figure~\ref{turnoff} shows
the cluster TO region in the CMD constructed using three different
color baselines\footnote{Note that, as mentioned above in the text,
the calibrated $U$ magnitude is not reliable at extreme colors because
of the large difference between the F336W and the Johnson $U$
bandpass. We illustrate it here because the relatively narrow width of
the main sequence in this plane more clearly indicates the $V$
magnitude of the turnoff point.}.  The horizontal line shows our best
estimate of the TO magnitude $V^{\rm TO}=17.65$, while the two dotted
lines define its error range estimate: $\pm 0.10$ magnitude. Coupled
with our distance determination, this gives a TO absolute magnitude
$M_V^{\rm TO}=17.65-13.27=4.38\pm0.17$.

There is little scatter among recent determination of the metallicity
of 47~Tuc, and we adopt [Fe/H]=--0.70 (Carretta \& Gratton 1997). It
is now generally believed that for metal poor halo stars and globular
clusters the $\alpha$--elements enhancement requires values of
[$\alpha$/Fe] between 0.4 and 0.6. However, following Gratton, Quarta
\& Ortolani (1986) we adopt [$\alpha$/Fe]$\approx 0.3$ for the
abundance ratios in 47~Tuc.

\begin{figure*}
\vskip 0.4cm
\centerline{\psfig{file=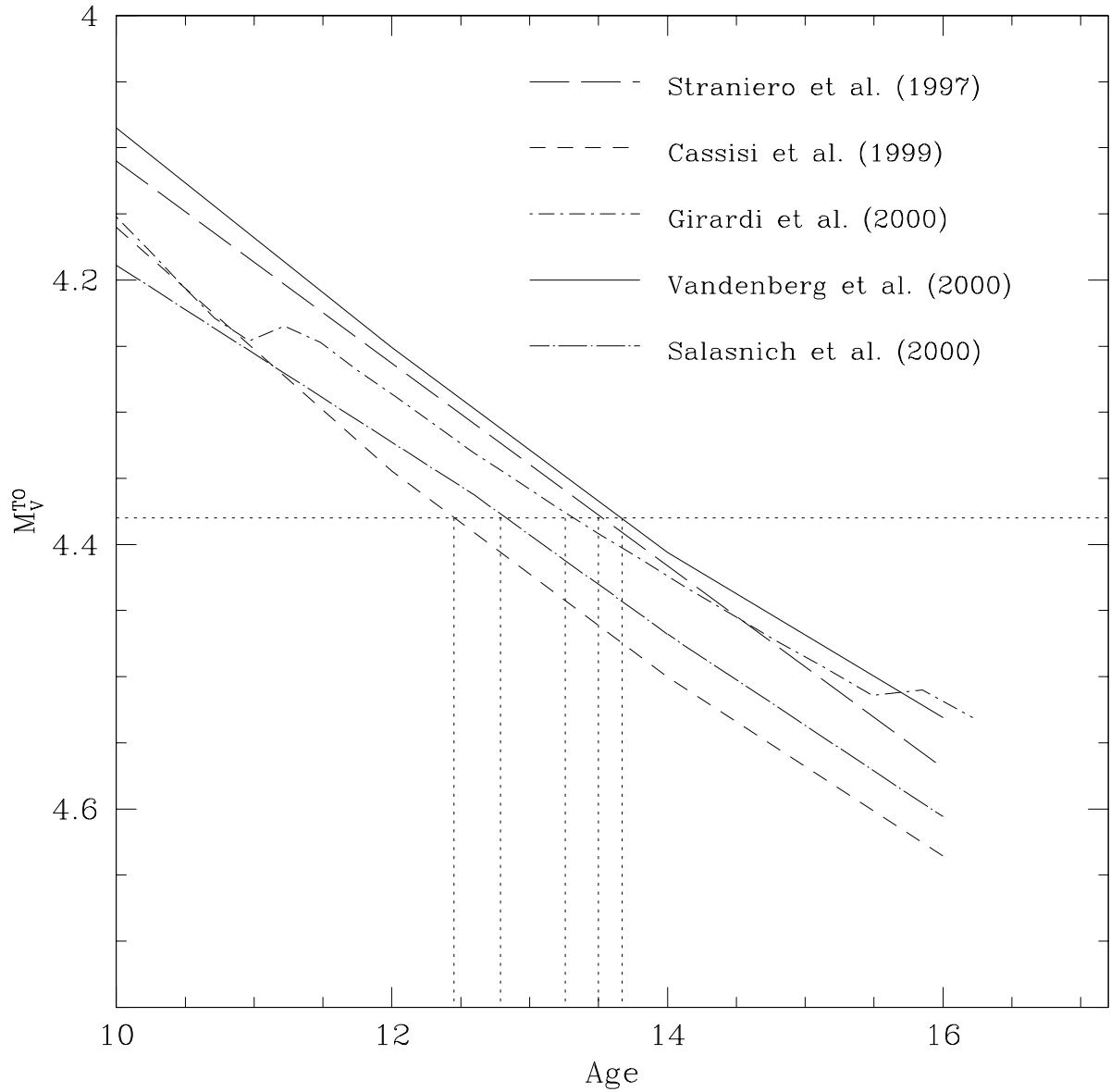,angle=0,width=17truecm}}
\caption[]{$M_V^{\rm TO}$ vs age relation derived from different
theoretical models. The horizontal dotted line shows the estimated
location of $M_V^{\rm TO}$ for 47~Tuc, and the vertical ones the
corresponding ages, according to the three models.}
\label{to_age}
\end{figure*}

Figure~\ref{to_age} shows the $M_V^{TO}$ vs age relation, for the
adopted metallicity [M/H]=--0.5, from the most recent models available
in the literature, and the values we derive are listed in Table~3. The
first column in Table~3 gives the reference of the adopted model, the
second and third give its metallicity and helium abundance, the fourth
indicates whether the helium and heavy element diffusion was included
in the models, and the next one gives the age corresponding to
47~Tuc. The average age obtained from the models that include atomic
diffusion is 12.9 Gyr, while the others give 13.5. As expected (e.g.,
Castellani et al.\ 1997) the two values differ by about 0.6 Gyr. Note
however that the ~1 Gyr discrepancy between the age from Cassisi et
al.\ (1999) and Straniero et al.\ (1997) is mainly due to the use of a
different equation of state in the two sets of models (OPAL the
former, and a modified version of the Straniero 1988 the latter; see
Cassisi et al.\ 1999, their Table~1). The age error corresponding to
the uncertainty in $M_V^{TO}$ is $\sim 2.2$ Gyr.

The last column of Table~3 shows the ages derived for NGC~6752
assuming $M_V^{\rm TO}=17.4-13.27$. Here $V^{\rm TO}=17.4$ (Penny \&
Dickens 1986) is the same value adopted in Paper I, while the distance
modulus $(m-M)_V=13.27$ comes from the value found in Paper~I
($(m-M)_V=13.17$) corrected for the fact that in Paper~I a thin
hydrogen envelope was assumed for the disk WDs, and therefore a
correction of $2.4\times 0.04$ is needed (see Sec.~2.2). As Table~1
clearly shows, the two clusters turn out to be coeval within the
errors. A forthcoming paper will be devoted to a more direct
comparison between the 47~Tuc and NGC~6752 turnoff locations, matching
the cluster WD sequences .


\section{Conclusions}

The distance of the Galactic globular cluster 47~Tuc as been measured
by comparing its WD cooling sequence to the fiducial sequence obtained
from a sample of local WDs of known trigonometric parallax.  We derive
an apparent distance modulus of $(m-M)_V=13.27\pm0.14$, or
$(m-M)_0=13.09\pm0.14$, implying a turnoff absolute magnitude of
$M_V^{\rm TO}=17.65-13.27=4.38\pm0.17$ and therefore an age of $\sim
13\pm2.2$ Gyr, if models including atomic diffusion are used.

We demonstrate again the feasibility of the method using deep optical
WFPC2 imaging, that permits very accurate measurement of magnitudes
even for stars as faint as the GC WDs. The low dispersion in the
cluster cooling sequence together with the fact that the method is not
affected by uncertainties in the metallicity, nor in the absolute
calibration of the photometry (instrumental magnitude system is
consistently used) permit us to measure the distance moduli of the
closest clusters with an error of 0.14 mag, which includes systematic
uncertainties.

The derived distance modulus is $\sim 0.42$ mag {\it shorter} than
an extreme value obtained with the subdwarf fitting method. Such
``long'' distance for 47~Tuc would imply an implausibly low mass
for the cluster WDs ($\sim 0.36 M_\odot$), or a very large systematic
error in the masses of the local WDs as obtained by Balmer-line
fitting.

Nevertheless, an important requirement for the application of this
method is the linearity of the detector at the two extremes of the
dynamical range. HST allows one to obtain very small {\it statistical}
photometric errors down to the faintest stars, but problems like the
CTE loss in the WFPC2 may affect the accuracy of the measures.
Therefore, the use of the appropriate correction for this effect can
be crucial for the result. We applied a state of the art determination
of the correction, but identify in the CTE problem the main possible
source of residual systematic bias.

However, to reconcile the ``long'' distance to 47~Tuc, the mismatch
between the cluster and the local WD sequences should be as high as
$\sim 0.36$ mag in U-I, which again we consider highly implausible.

%
%

\acknowledgments

We are deeply grateful to Ivan King for allowing us to measure its
proprietary frames from program GO8160. We also thank Peter Stetson
for providing us the mask files for vignetting and pixel area
correction for WFPC2, and the WFPC2 PSF models that made the data
reduction straightforward. We are grateful to Raffaele Gratton for 
useful discussions.

J.B.H. and R.M.R. wish to acknowledge the support for this work provided by
NASA through grant GO~7465 from the Space Telescope Science Institute,
which is operated by the Association of Universities for Research in
Astronomy, Inc., under NASA contract NAS5-26555.

FRF aknowleges the financial support \\
(MM02241491\_004) of the Ministero
della Ricerca Scientifica e Tecnologica to the project {\it Stellar 
Observable of Cosmological Relevance.}

%

%
%

\clearpage
\begin{deluxetable}{cccccc}
\tablewidth{0pt}
\tablenum{1}
\tablecaption{THE LOCAL CALIBRATING WHITE DWARFS}
\tablehead{
\colhead{WD} &
\colhead{$\pi\pm\Delta\pi$} &
\colhead{Ref.} &
\colhead{$M\pm\Delta M$} &
\colhead{Ref.} &
\colhead{WD type}
}
\startdata
WD0002+729  &  $0.0291\pm0.0047$  &  1  &  $0.600\pm0.030$  &  2 &  DB  \\
WD0644+375  &  $0.0626\pm0.0018$  &  1  &  $0.655\pm0.020$  &  3 &  DA  \\
WD1327-083  &  $0.0611\pm0.0028$  &  1  &  $0.502\pm0.017$  &  4 &  DA  \\
WD1917-077  &  $0.1010\pm0.0026$  &  1  &  $0.550\pm0.050$  &  5 &  DB  \\
WD1935+327  &  $0.0561\pm0.0029$  &  1  &  $0.512\pm0.013$  &  6 &  DA  \\
WD2126+734  &  $0.0433\pm0.0035$  &  1  &  $0.513\pm0.012$  &  6 &  DA  \\
WD2326+049  &  $0.0725\pm0.0048$  &  1  &  $0.690\pm0.025$  &  7 &  DA  \\
WD2341+322  &  $0.0559\pm0.0017$  &  1  &  $0.494\pm0.021$  &  6 &  DA  \\
\tableline
\enddata
\tablenotetext{}{
References: 
(1) Van Altena et al. (1991); 
(2) Beauchamp (1995); 
(3) Bergeron, Saffer \& Liebert (1992);
(4) Bragaglia et al. (1995);
(5) Oswalt et al. (1991);
(6) Bragaglia \& Bergeron (2000); 
(7) Bergeron et al. (1995) 
}
\end{deluxetable}
\begin{deluxetable}{lccl}
\tablewidth{0pt}
\tablenum{2}
\tablecaption{47~TUC DISTANCE DETERMINATIONS}
\tablehead{
\colhead{ Method } &
\colhead{ $(m-M)_V$ } &
\colhead{ $E(B-V)$ } &
\colhead{ Reference } \\
}
\startdata
HB fitting       & $13.40\pm0.20$ & 0.040 & Hesser et al. (1987)    \\
Baade-Wesselink  & $13.36\pm0.17$ & 0.040 & Storm et al. (1994)     \\
HB fitting       & $13.45\pm0.07$ & 0.040 & Kaluzny et al. (1998)   \\
HB fitting       & $13.50\pm0.05$ & 0.050 & Salaris \& Weiss (1998) \\
HB fitting       & $13.43\pm0.15$ & 0.040 & Ferraro et al. (1999)   \\
RGB tip		 & $13.42\pm0.20$ & 0.040 & Ferraro et al. (2000)   \\
MS fitting       & $13.69\pm0.15$ & 0.040 & Reid (1998)             \\
MS fitting       & $13.55\pm0.09$ & 0.055 & Carretta et al (2000)   \\
WDs              & $13.27\pm0.14$ & 0.055 & This work               \\
\tableline
\enddata
\end{deluxetable}
\begin{deluxetable}{lcccll}
\tablewidth{0pt}
\tablenum{3}
\tablecaption{THE AGE OF 47~TUC USING DIFFERENT MODELS}
\tablehead{
\colhead{Model} &
\colhead{[M/H]} &
\colhead{Y} &
\colhead{diffusion} &
\colhead{age (Gyr)} &
\colhead{NGC~6752 age}
}
\startdata
Straniero et al. (1997)    & $-$0.60 & 0.230 & yes  & 13.5 & 13.0 \\
Cassisi   et al. (1999)    & $-$0.52 & 0.230 & yes  & 12.4 & 11.7 \\
VandenBerg et al. (2000)   & $-$0.62 & 0.241 & no   & 13.7 & 13.3 \\
Girardi    et al. (2000)   & $-$0.50 & 0.245 & no   & 13.3 & 13.5 \\
Salasnich et al. (2000)    & $-$0.50 & 0.245 & yes  & 12.8 &      \\
\tableline
\enddata
\end{deluxetable}

\newpage
\newpage
\newpage
\newpage
\newpage
\newpage
\newpage
\newpage
\newpage

\end{document}